# China's Shrinking Home Bias and Rising Disruptive Impact: Evidence from a Global Citation Network Analysis

Nan Deng[*], Zheng Ma[**], Xiaohui Liu[***]

[*]*dengnan@mail.las.ac.cn*
https://orcid.org/0000-0001-6846-941X
National Science Library, Chinese Academy of Sciences, Beijing 100190, China.
Department of Information Resources Management, School of Economics and Management, University of Chinese Academy of Sciences, Beijing 100190, China.

[**]*mazheng@mail.las.ac.cn*
https://orcid.org/0000-0002-9000-1327
National Science Library, Chinese Academy of Sciences, Beijing 100190, China.

[***]*liuxhui612@163.com*
https://orcid.org/0000-0002-7233-7820
Shandong Normal University Library, Shandong 250358, China

China has become the world's largest producer of scientific publications, yet concerns persist that this growth is inflated by excessive domestic citation practices. In this study, we analyze a citation network of over 45 million publications from Web of Science (1980–2025) to investigate China's home citation bias and research impact. Using a network reshuffling null model to control for the structural effect of publication volume, we find that China's home citation bias is less pronounced than commonly assumed and has been steadily declining over the past two decades. Chinese researchers do not exhibit a significantly stronger home citation preference than other major countries, indicating increasing internationalization rather than insularity. Furthermore, using the persistent disruption framework, we show that Chinese papers are converging toward American papers in their capacity to produce paradigm-shifting work. These findings challenge prevailing narratives about Chinese scientific home bias and suggest that China's  advances in research impact.

## 1. Introduction

In recent years, China's academic influence has expanded rapidly, positioning the country as a global leader in scientific output (Xie & Freeman, 2019). According to statistics from the U.S. National Science Foundation (NSF) based on the Scopus database, China surpassed the United States in 2016 to become the world's largest producer of scientific research (Tollefson, 2018). By 2018, the number of research papers from China indexed in SCI also exceeded that of the United States (Pisani et al., 2025). China has also overtaken the United States in the number of degrees awarded at various levels (Zhu & Liu, 2020), and its leadership in international collaborations, including Sino-U.S. cooperation, continues to grow annually (Wu et al., 2025).

Beyond sheer volume, China's research quality has also improved markedly. In recent years, China's contribution to the world's top 1% of high-impact research has caught up with or even surpassed that of the United States, particularly in engineering, electronics, materials science, physics, and chemistry (Xie et al., 2014; Springer Nature, 2023). The journal Science has noted that both submissions from China and their acceptance rates are increasing (Thorp, 2024).

Despite these achievements, China's rapid ascent has raised questions about whether its scientific influence is genuine or inflated by domestic citation practices (Normile, 2024). A study released by Japan's National Institute of Science and Technology Policy (NISTEP) reported that 62% of citations for the top 10% most-cited Chinese articles originate from within China, compared to only 24% for the United States (NISTEP, 2022). Furthermore, when national self-citations are excluded, China's ranking in top-journal influence (2000–2021) drops from second to fourth place, behind the United States, the United Kingdom, and Germany (Qiu et al., 2025a). These findings have fueled skepticism about the true strength of Chinese science.

The phenomenon whereby researchers disproportionately cite articles from their own country is known as home bias. Research on home bias dates back to 1985 (Lange, 1985) and has been extensively studied in finance and economics, where investors exhibit strong preferences for domestic assets (French & Poterba, 1991; Coval & Moskowitz, 1999; Ahearne et al., 2004; Van Nieuwerburgh & Veldkamp, 2009). In scientometrics, similar patterns have been documented: research papers are read significantly more frequently in their country of origin (Thelwall & Maflahi, 2015), domestic citation rates are elevated across all countries (Larivière et al., 2018), and this tendency persists even after controlling for individual-level self-citations (Khelfaoui et al., 2020). Moreover, self-citation rates have shown an increasing trend over time (Baccini et al., 2019).

Compared to U.S. papers, highly cited Chinese papers exhibit a higher proportion of domestic citations (Tang et al., 2015). However, less-cited papers tend to be more grounded in domestic research, while highly cited papers draw more heavily on international work (Bornmann et al., 2018). National citation patterns are also more pronounced in the social sciences and humanities than in the natural sciences (Khelfaoui et al., 2020). Beyond home bias, cross-country citation disparities exist: Chinese scientists' papers receive fewer citations from U.S. researchers compared to papers from other countries, even after controlling for quality (Qiu et al., 2025b), and papers not published in U.S. journals struggle to obtain citations from the American scientific community (Sjöberg, 2000). Conversely, all countries exhibit a tendency to over-cite U.S. papers, reflecting the dominant status of the United States in global science (Larivière et al., 2018).

Various factors shape national self-citation rates. National policies centered on citation metrics can lead to abnormal increases in home bias, as exemplified by Italy (Baccini et al., 2019; Baccini & Petrovich, 2023). Research output volume also plays a critical role: as a country's scientific production increases, it occupies a larger share of the citation space, mechanically raising its self-citation rate. One study demonstrated that self-citation rates increase with the logarithm of output (Larivière et al., 2018).Other explanations for elevated self-citation rates in developing countries include a focus on applied science addressing national development needs, weak norms for citing references, inadequate graduate training, a preference for native-language literature, and the proliferation of low-quality domestic journals (Ladle et al., 2012). The widening gap between international and domestic research may also affect countries' ability to draw from the global knowledge base (Adams, 2013).

Methodologically, national-level citation analysis distinguishes between the perspective of referencing (home reference rate) and the perspective of being cited (home citing rate) (Shehatta & Al-Rubaish, 2019; Lawani, 1982; Glänzel & Schubert, 2005). The home reference rate depends solely on the citation behavior of a country's own scholars, whereas the home citing rate is also influenced by how researchers in other countries cite that nation's work (Khelfaoui et al., 2020). In contrast to self-citation rates, foreign citation rates are often interpreted as measures of international influence (Hassan & Haddawy, 2013).

Despite the extensive literature on home bias, a critical question remains underexplored: to what extent is China's elevated home citation rate attributable to genuine preferential citation behavior versus the mechanical effect of its rapidly expanding publication volume? Previous studies have documented China's high domestic citation rates but have rarely disentangled these two contributing factors.

In this study, we address this gap by constructing a citation network from the Web of Science database spanning 1980 to 2025 and employing a network reshuffling null model to establish expected citation patterns under random conditions. By comparing observed home reference and home citation rates against this null baseline, we separate the structural effect of publication volume from citation preferences. We further extend our analysis beyond citation volume to examine research quality using the persistent disruption framework (Deng et al., 2025). This two-dimensional framework assesses whether publications contribute original, paradigm-shifting ideas that maintain their foundational status over time. By comparing persistent disruption across major countries, we show that China has been converging toward Western countries in disruptive research output.

## 2. Results

As shown in Figure 1a, China's scientific publication output has grown steadily over the past several years, making it the largest source country for research publications worldwide. The growth shows no sign of deceleration, and the gap between China and other major research nations, including the USA, continues to widen. In terms of publications in top journals (Nature, Science, and Cell), the USA still leads globally, but China has surpassed the UK to become the second-largest contributor (Figure 1b). In addition, the citation impact of Chinese publications has improved notably (Figure 1c). We used the category-normalized citation impact (CNCI) to measure publication influence, controlling for differences in field and publication year.

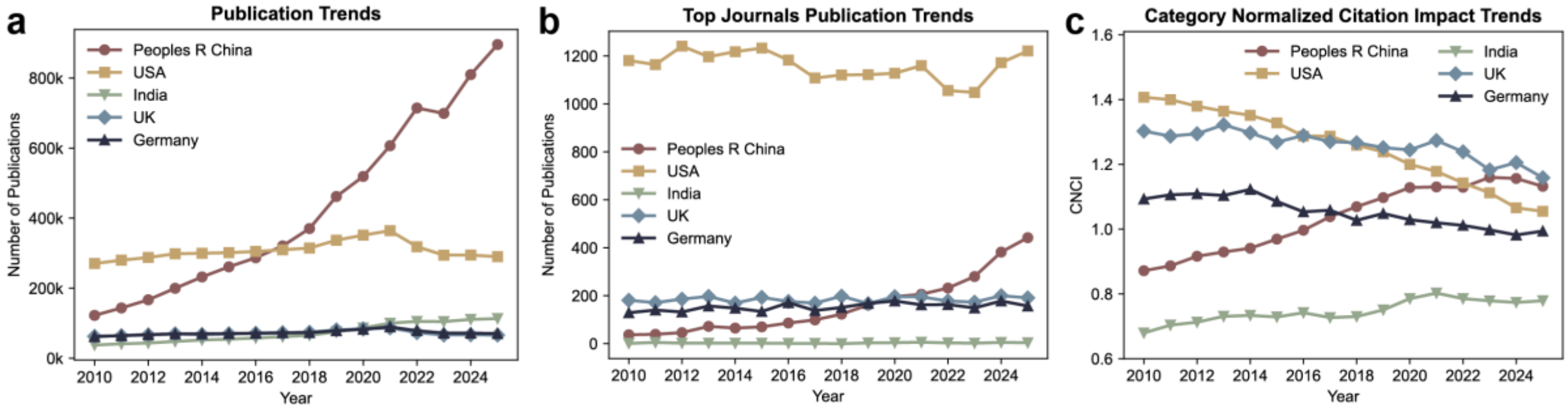

Figure 1: (a) Publication output, (b) top-journal publications, and (c) citation impact trends of 5 major countries.

We evaluate a country's home bias using two indicators: home reference rate and home citation rate (see Methods). China's home reference rate has risen notably over the past decade, surpassing other major countries and approaching the level of the USA (Figure 2a). In terms of citations received, China has maintained the highest home citation rate among all major countries, consistently exceeding 50% and continuing to rise (Figure 2b). These findings suggest that Chinese researchers increasingly cite domestic publications, while Chinese publications also attract a large share of citations from within China.

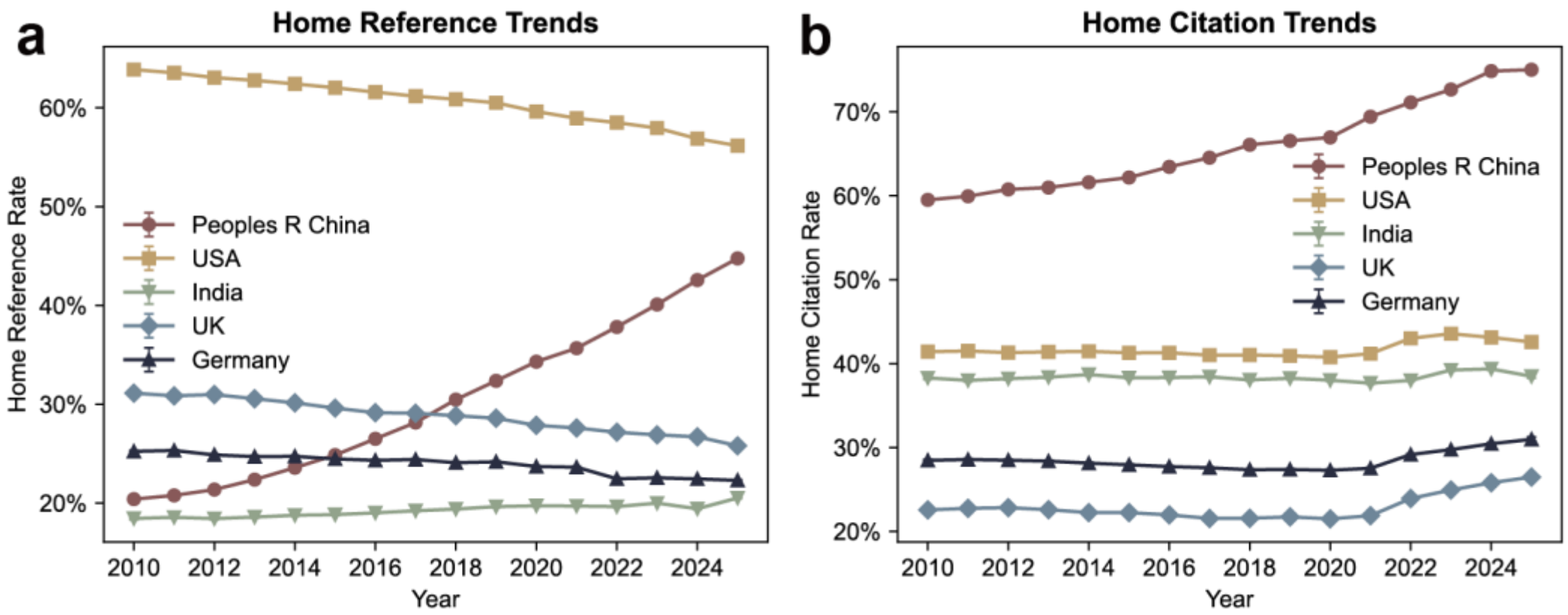

Figure 2: (a) Home reference rate and (b) home citing rate of 5 major countries.

To examine whether Chinese science exhibits a strong home bias, that is, whether Chinese researchers disproportionately cite domestic publications beyond what publication scale alone would predict, we constructed a null model by randomizing the citation network (see Methods). The randomization preserves key structural constraints, including the publication year of focal papers and their references', as well as the degree distribution of the network. This procedure generates the expected home reference and home citation rates under random citation behavior. We then compare these baselines with the observed rates.

Figures 3a and 3b present the expected home reference rate and home citation rate under the null model for the five major countries. The USA shows a high expected home reference rate due to its large publication volume. For China, the expected home reference rate increased rapidly after 2010, reflecting the growth in Chinese publications. A similar trend is observed for the expected home citation rate, where China ranks among the highest (Figure 3b). These expected trends largely mirror the observed patterns in Figure 2, suggesting that much of the increase in China's home reference and home citation rates can be attributed to the expanding volume of Chinese output rather than preferential citation behavior.

To separate the structural effect of publication volume from citation preferences, we calculated the ratio of observed rates to expected rates. Figure 3c shows a notable finding: China's relative home reference tendency (the observed-to-expected, O/E, ratio) has declined steadily over the past decade and a half. Crucially, this ratio has remained above 1 throughout, indicating that—relative to a null model of random citation—Chinese researchers still cite domestic work more often than expected by chance; however, this home bias has been weakening over time, converging toward the random baseline. This is notable because the rapid growth of Chinese publications would, under random citation behavior, lead one to expect an increasing share of domestic citations; instead, the propensity to over-cite domestic work is diminishing. In contrast, the relative home reference tendencies of the USA, India, the UK, and Germany have remained stable.

A similar pattern is observed for the home citation rate. As shown in Figure 3d, China's relative home citation ratio is comparatively low among the five countries and has been declining, indicating that Chinese publications receive a smaller share of domestic citations than other major countries relative to random expectation. The ratio remains above 1, meaning Chinese papers still receive more domestic citations than the null model would predict, but the declining trend suggests that China's home bias is shrinking rather than intensifying.

These findings challenge the prevailing view that Chinese science is characterized by excessive self-citation. Our analysis shows that China's home citation bias is lower than commonly assumed and is actively diminishing. This pattern may reflect the increasing internationalization of Chinese research and a growing emphasis on citing high-quality work regardless of national origin. Overall, these results suggest that Chinese science is becoming progressively more integrated into the global knowledge network.

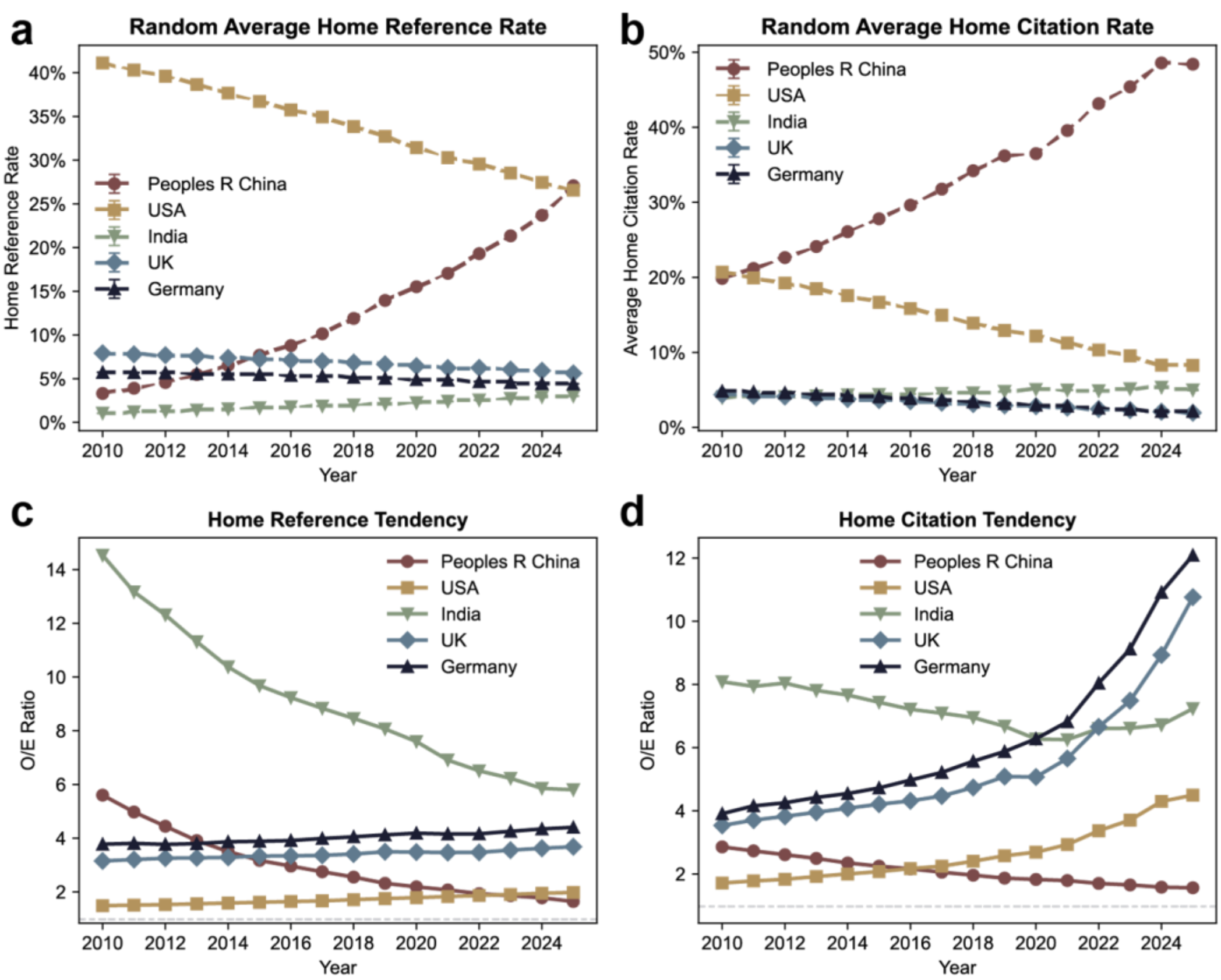

Figure 3: Expected home reference and home citation rates under the null model (a, b) and the ratio of observed to expected rates (c, d) for the major countries.

In addition to citation impact, we examine whether Chinese research is disrupting existing scientific paradigms. We employ the persistent disruption framework to compare the persistent disruption (see Methods) of publications from the USA and China. This two-dimensional framework captures both reference disruption (the extent to which a paper departs from its cited references) and citation disruption (the extent to which a paper influences its citing papers in a consolidating or disruptive manner).

Figure 4a shows that China's reference disruption has steadily increased since 2000, converging with that of Western countries. Although Chinese papers still lag behind other major nations in citation disruption, this gap is actively narrowing (Figure 4b). Figure 4c integrates both dimensions, with dot opacity representing temporal progression, revealing that China’s overall persistent disruption is rapidly approaching Western levels. This dual trend highlights that China is not only producing highly disruptive research but also gaining commensurate academic recognition. This convergence suggests that China has not only expanded its publication volume but has also improved the innovation of its research output. The narrowing gap in persistent disruption indicates that Chinese science is increasingly contributing original ideas rather than merely consolidating existing knowledge. These findings complement the citation-based analyses presented earlier, suggesting that China's rise in global science extends beyond quantity to qualitative dimensions of scientific impact.

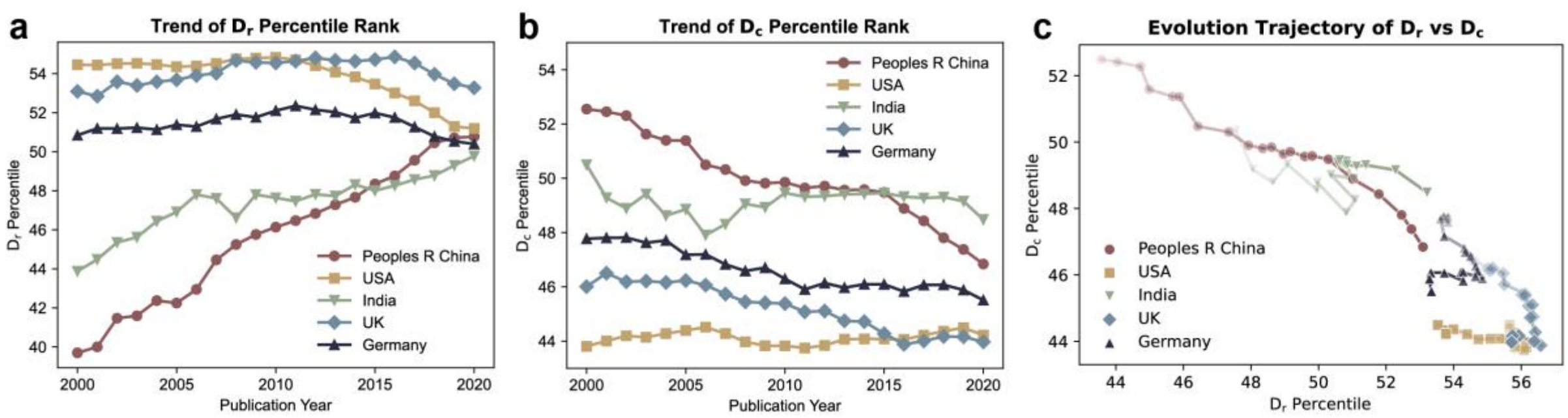

Figure 4: Persistent disruption of major countries.

## 3. Discussion

This study set out to test a widely held assumption that China's rapid ascent in scientific impact has been inflated by an unusually strong tendency to cite domestic work. Our analysis runs counter to this assumption. After using a network reshuffling null model to hold publication volume effect constant, we find that China's home citation bias is more modest than commonly portrayed and has been declining over the period studied. This matters because much of the existing commentary draws its conclusions directly from China's high observed rate of domestic citation, without accounting for the fact that, as a national corpus grows, the probability of citing domestic work rises even under entirely random citation behavior. Once this baseline is accounted for, much of the apparent home bias proves to be an artifact of scale.

All countries show a relative home bias above the random baseline. In other words, researchers everywhere cite domestic work more often than chance would predict, so home bias is a general feature of science rather than something unique to China. China’s home bias is no stronger than that of other major producers. In fact, China's home bias has been steadily declining over time. We also concerns research impact. Applying the persistent disruption framework, we find that Chinese papers are increasingly capable of producing genuinely paradigm-shifting work.

Several limitations qualify our conclusions. First, each paper is attributed to a single country based on the first author's primary affiliation, which does not reflect the collaborative reality of modern science. Second, a high home citation rate need not signal deliberate preference. When a country leads a research field, much of the most relevant literature is legitimately its own, and citing it reflects citing the best available work rather than bias. Because our null model controls for publication volume but not for field-level citation norms, part of the home bias we measure may stem from this kind of warranted concentration of expertise rather than from genuine preference.

## 4. Method

### *4.1. Data*

This study uses bibliometric data from the Web of Science (WoS) Core Collection, covering the period from 1980 to 2025. The dataset includes articles and reviews indexed in WoS. After removing records with incomplete metadata (e.g., missing author information, publication year, or reference lists), we constructed a directed citation network comprising over 44 million nodes (publications) and over 1.8 billion edges (citation links). We also extracted metadata for each publication, including publication year, journal name, author affiliations, and country or region. The country of each publication was assigned based on the first author's institutional affiliation, meaning each paper is attributed to a single country.

*4.2. Home Reference & Citation Rate*

For each paper $i$, we assign a single country of origin $c$ based on the affiliation of its first author's first listed institution. We then define two complementary indicators of domestic orientation, each a proportion bounded in $[0,1]$. The *Home Reference Rate* measures the extent to which a paper draws on domestic prior work. It is the share of $i$'s cited references that originate from country $c$:

$$HRR_i = \frac{R_i^c}{R_i}$$

where $R_i$ is the number of references in paper $i$ with an assignable country of origin, and $R_i^c$ is the number of those references attributed to country $c$.

The *Home Citation Rate* measures the extent to which a paper is taken up by domestic subsequent work. It is the share of $i$'s incoming citations that come from country $c$:

$$HCR_i = \frac{C_i^c}{C_i}$$

where $C_i$ is the number of citing papers with an assignable country of origin and $C_i^c$ is the number of those citing papers from country $c$. Note that both rates are defined as proportions in $[0,1]$ and are therefore distinct from the conventional notion of a citation rate as citations received per paper.

*4.3. Network Reshuffle*

To assess whether the observed citation patterns deviate from random expectation, we employ a reshuffling method widely used in network science (Uzzi et al., 2013). Specifically, we construct a null model by randomly rewiring the citation network while preserving key structural constraints. For each focal paper, we fix the publication years of both the focal paper and its references, then randomly reshuffle the connections between them. This procedure preserves the out-degree (number of references) of each focal paper and the in-degree (number of citations received) of each referenced paper. The reshuffling is repeated 10 times to generate a distribution of expected values under random citation behavior. By comparing the observed patterns against this null distribution, we identify statistically significant deviations that reflect non-random citation preferences, allowing us to control for the structural effect of publication volume and isolate country-level citation tendencies.

*4.4. Persistent Disruption*

To assess the qualitative dimension of scientific impact, we employ the concept of persistent disruption recently introduced by Deng et al. (Deng et al., 2025). The original disruption index, proposed by Funk and Owen-Smith (Funk & Owen-Smith, 2017) and subsequently applied to scientific publications by Wu et al. (Wu et al., 2019), measures the extent to which a paper renders its references obsolete by examining whether citing papers also cite the focal paper's references. However, this traditional metric has notable limitations: highly disruptive papers are not necessarily milestone works and may even receive few citations (Park et al., 2023). To address this issue, Deng et al. developed a link disruption metric that quantifies the disruptiveness of each individual citation link.

For a focal paper $p$ and one of its references $r$, the link disruption $d_{pr}$ is defined as:

$$d_{pr} = \frac{n_p - n_{pr}}{n_p + n_{pr} + n_r}$$

where $n_p$ denotes the number of papers that cite the focal paper $p$ but not the reference $r$, $n_{pr}$ denotes the number of papers that cite both $p$ and $r$, and $n_r$ denotes the number of papers that cite the reference $r$ but not the focal paper $p$. This metric ranges from $-1$ (fully consolidating) to 1 (fully disruptive).

Based on this link-level metric, the reference disruption $D_r$ of a focal paper $p$ is calculated as the average link disruption across all its references:

$$D_r = \frac{1}{|R_p|} \sum_{r \in R_p} d_{pr}$$

where $R_p$ is the set of references of paper $p$. A higher $D_r$ indicates that the focal paper more strongly disrupts its cited references.

Similarly, the citation disruption $D_c$ measures the extent to which the focal paper $p$ is disrupted by its citing papers. The citation disruption $D_c$ is then calculated as:

$$D_c = \frac{1}{|C_p|} \sum_{r \in C_p} d_{cp}$$

where $C_p$ is the set of citing papers of paper $p$. A lower $D_c$ (more negative) indicates that the focal paper is more consolidated by its citing papers and thus retains its foundational status.

This two-dimensional framework enables the identification of "persistently disruptive" papers—those that exhibit high reference disruption (large $D_r$) but low citation disruption (small $D_c$). Such papers not only break from prior work but also maintain their foundational status without being superseded by subsequent research. Empirical analysis demonstrates that persistently disruptive papers are more likely to be recognized as award-winning works, including Nobel Prize-winning papers, and receive substantially higher citations.

**Open science practices**

The bibliometric data used in this study were obtained from the Web of Science (WoS) Core Collection under an institutional license. Due to the terms of the licensing agreement, the raw data cannot be publicly shared.

**Acknowledgments**

This research was supported by the project of the national S&T innovation strategic research (2024GH055) for the 15th Five-Year plan of China. And This work is also supported by Shandong Provincial Natural Science Foundation (Project Number: ZR2023QG129).

**Author contributions**

N.D and Z.M. designed the research, N.D. performed the experiments, N.D. and Z.M. analyzed the data and all authors wrote the paper.

**Competing interests**

The authors declare no competing interests.